\documentclass{elsart1p}

\usepackage{amssymb}
\usepackage{latexsym} 

\newcommand{\p}{\partial}
\newcommand\eqn[1]{(\ref{#1})}      
\newcommand\Eqn[1]{Eq.~(\ref{#1})}  
\newcommand\Ref[1]{Ref.~\cite{#1}}  
\newcommand{\beq}{\begin{equation}}
\newcommand{\eeq}{\end{equation}}
\newcommand{\ba}{\begin{array}}
\newcommand{\bea}{\begin{eqnarray}}
\newcommand{\ea}{\end{array}}
\newcommand{\eea}{\end{eqnarray}}
\newcommand{\nn}{\nonumber\\}

\newcommand{\tr}{{\rm  tr}}

\newcommand{\Str}{\hbox{Str}}
\newcommand{\Ln}{\hbox{Ln}}
\newcommand{\inter}{{\mbox{\scriptsize int}}}
\newcommand{\Gammatpi}{\Gamma_{\mbox{\scriptsize 2PI}}}
\newcommand{\Gammaint}{\Gamma_\inter}
\newcommand{\bcG}{{\bar{\cal G}}}

\newcommand{\cG}{{\cal G}}

\def\slashchar#1{\setbox0=\hbox{$#1$}           
   \dimen0=\wd0                                 
   \setbox1=\hbox{/} \dimen1=\wd1               
   \ifdim\dimen0>\dimen1                        
      \rlap{\hbox to \dimen0{\hfil/\hfil}}      
      #1                                        
   \else                                        
      \rlap{\hbox to \dimen1{\hfil$#1$\hfil}}   
      /                                         
   \fi}

\begin{document}

\begin{frontmatter}



\title{2PI functional techniques for abelian gauge theories}


\author[Urko]{U. Reinosa}
 and 
\author[Julien]{J. Serreau}

\address[Urko]{Centre de Physique Th\'eorique, \'Ecole Polytechnique, 91128 Palaiseau, France\thanksref{CPhT}}
\thanks[CPhT]{CPHT is unit\'e mixte de recherche UMR7644 (CNRS, \'Ecole Polytechnique).}
\address[Julien]{Astro-Particule et Cosmologie,
Universit\'e Paris 7 - Denis Diderot,\\
10, rue A. Domon et L. Duquet, 75205 Paris Cedex 13, France\thanksref{APC}}
\thanks[APC]{APC is unit\'e mixte de
recherche UMR7164 (CNRS, Universit\'e Paris 7, CEA, Observatoire de Paris).}

\begin{abstract}
We summarize our recent work \cite{Reinosa:2007vi,Reinosa:2006cm,QED2} concerning the formulation of two-particle-irreducible (2PI) functional techniques for abelian gauge field theories. 
\end{abstract}

\end{frontmatter}

\section{Introduction}

Two-particle-irreducible (2PI) functional techniques provide a powerful tool to devise systematic nonperturbative approximation schemes in quantum field theory, of interest in numerous physical situations where standard (loop, $1/N$) expansions fail. A non-trivial issue for such approximations is that they should reflect, as much as possible, basic features of the exact theory, such as symmetries and renormalizability.

The issue of renormalization in the 2PI formalism has attracted a lot of attention in recent years \cite{vanHees:2001ik,Berges:2005hc,Berges:2004hn,Cooper:2004rs,Reinosa:2005pj} and is now a well established topic. The basic tools have been put forward in pioneering works \cite{vanHees:2001ik} and a complete 2PI renormalization theory has been proposed  for theories with linearly realized global symmetries \cite{Berges:2005hc}. 

In a recent series of papers \cite{Reinosa:2007vi,Reinosa:2006cm,QED2}, we have developed the 2PI renormalization theory for abelian gauge theories. This requires a proper understanding of symmetry constraints, i.e. Ward-Takahashi (WT) identities, in the 2PI formalism, which is the purpose of \Ref{Reinosa:2007vi}. In Refs.~\cite{Reinosa:2006cm,QED2}, these results are used to develop a consistent renormalization procedure for 2PI QED, which preserves the underlying gauge-symmetry. The present contribution provides a summarized account of these papers.

\section{2PI effective action for QED}

We consider QED in the covariant gauge and use dimensional regularization. The gauge-fixed classical action reads, with standard notations,
\begin{equation}
\label{eq:classact}
S[A,\psi,\bar\psi]=\int d^dx \left\{\bar\psi\Big[i\slashchar{\partial}-e\slashchar{A}-m\Big]\psi+\frac{1}{2}A^\mu\Big[g_{\mu\nu}\partial^2-(1-\lambda)\partial_\mu\partial_{\nu}\Big]A^\nu\right\}\,,
\end{equation}
where $\lambda$ is the gauge-fixing parameter. Aside from the gauge-fixing term, the classical action is invariant under the gauge transformation
\begin{equation}\label{eq:gauge}
\psi(x)\rightarrow e^{i\alpha(x)}\psi(x)\,,\;\;
\bar\psi(x)\rightarrow e^{-i\alpha(x)}\bar\psi(x)\,,\;\;
A_\mu(x)\rightarrow A_\mu(x)-\frac{1}{e}\,\partial_\mu\alpha(x)\,,\nonumber\\
\end{equation}
where $\alpha(x)$ is an arbitrary real function. To define the 2PI effective action, it is convenient to grab the bosonic and fermionic connected one- and two-point functions in a superfield $\varphi$ and corresponding supercorrelator $\cG$ (transposition includes space-time variables):  
\bea
\label{eq:sfield}
 \varphi=\left(
 \begin{tabular}{c}
  $A$\\
  $\psi$\\
  $\bar\psi^t$
  \end{tabular}
  \right),\qquad 
  \cG=\left(
 \begin{array}{ccc}
 G&K^t&\bar K\\
 K&F&D\\
 \bar K^t&-D^t&\bar F
 \end{array}
 \right)\,.
\eea
In general, one should keep track, at the level of the 2PI effective action, of all possible fields and correlators. Besides the usual photon and fermion two-point functions $G$ and $D$, this includes all possible mixed correlators, such as, for instance the photon-fermion correlators $K$ and $\bar K$, etc. Although the latter vanish on-shell, they should be set to their physical value only after all the relevant functional derivatives have been taken.\footnote{For instance, $\delta \Gammatpi/\delta K\delta\bar K$ does not vanish on-shell.}

Writing the classical action as $\smash{S[\varphi]=S_0[\varphi]+S_{\rm int}[\varphi]}$, where $\smash{S_0[\varphi]=\frac{1}{2}\varphi_m i\cG_{0,mn}^{-1}\,\varphi_n}$ is the quadratic part, with $\cG_0^{-1}$ the free inverse (super)propagator, the 2PI functional can be pa\-ra\-met\-ri\-zed as \cite{Reinosa:2007vi,Calzetta:2004sh}:
\beq 
\label{eq:2PI} \Gammatpi[\varphi,\cG]=S_0[\varphi]+\frac{i}{2}\,\Str\,\Ln\cG^{-1}+\frac{i}{2}\,\Str\,\cG_0^{-1}\cG+\Gammaint[\varphi,\cG]\,,
\eeq
where $\Str$ denotes the functional supertrace and $i\Gamma_{\rm int}[\varphi,\cG]$ is the sum of closed
two-particle-irreducible (2PI) diagrams made of classical QED vertices and lines given by~$\cG$.

The physical correlator $\bcG[\varphi]$ in the presence of a nonvanishing field $\varphi$ is obtained as the solution of the stationarity condition $\smash{\delta\Gammatpi/\delta \cG=0}$:\footnote{Here, $(-1)^{q_n}=1$ if $n$ refers to a bosonic superfield component and $(-1)^{q_n}=-1$ otherwise.}.
\beq
\label{eq:stat}
 \bcG^{-1}[\varphi]=\cG^{-1}_0-\bar\Sigma[\varphi]\quad{\rm with}\quad\bar\Sigma_{mn}[\varphi]\equiv \left.(-1)^{q_n}\,2i\,\frac{\delta\Gammaint[\varphi,\cG]}{\delta \cG_{nm}}\right|_{\bcG[\varphi]}\,,
\eeq

Finally, the effective action $\Gamma[\varphi]$, the generating functional for 1PI $n$-point vertex functions, is obtained as 
\beq
\label{eq:1PI}
 \Gamma[\varphi]\equiv\Gammatpi[\varphi,\bcG[\varphi]]\,.
\eeq
This is a mere identity in the exact theory. However, for finite 2PI approximations, Eqs~\eqn{eq:2PI}-\eqn{eq:1PI} define a powerful systematic resummation scheme for the effective action \cite{Berges:2005hc,QED2}. 

\section{Vertex functions}

In the 2PI framework, $n$-point vertex functions can be obtained in different ways, see e.g. \cite{Berges:2005hc}. Equivalent in the exact
theory, they differ in general at finite approximation order. The most straightforward definition involves the $n$-th derivatives of the (2PI-resummed) effective action \eqn{eq:1PI}:
\beq
\label{eq:npoint1}
 \Gamma^{(n)}_{1\ldots n}\equiv\left.\frac{\delta^n\Gamma[\varphi]}{\delta\varphi_n\cdots\delta\varphi_1}\right|_{\bar\varphi}\,,
\eeq
taken at the physical value $\bar\varphi$ of the field, defined by $\smash{\delta\Gamma[\varphi]/\delta\varphi|_{\bar\varphi}=0}$. We refer to these as 2PI-resummed vertex functions.

Other possible definitions of vertex functions involve derivatives of the 2PI effective action \eqn{eq:2PI} with respect to $\cG$. For instance, the two-point function -- the self-energy -- can either be obtained from the second derivative of the 2PI-resummed effective action, see \Eqn{eq:npoint1}, or directly from \Eqn{eq:stat}. In turn, higher $n$-point functions can be obtained as field-derivatives of the self-energy $\bar\Sigma[\varphi]$:
\beq
\label{eq:npoint2}
 iV^{(p+2)}_{mn;1\cdots p}\equiv\left.(-1)^{q_m}\frac{\delta^p\bar\Sigma_{nm}[\varphi]}{\delta\varphi_{p}\cdots\delta\varphi_{1}}\right|_{\bar\varphi}\,.
\eeq
Of course, $\smash{\Gamma^{(n)}=V^{(n)}}$ in the exact theory \cite{Reinosa:2007vi}, but this is not true in general at finite approximation order. We refer to the vertices \eqn{eq:npoint2} as 2PI vertex functions\footnote{We stress though that these are really proper, one-particle-irreducible vertex functions.}. Together with 2PI-resummed vertices \eqn{eq:npoint1}, they play a crucial role in the renormalization program.

\section{Symmetries and 2PI Ward-Takahashi identities \cite{Reinosa:2007vi}}

To analyze the role played by the gauge symmetry in the quantum theory, we write
\beq
\label{eq:symaction}
 S[\varphi]=S_{\rm sym}[\varphi]+S_{\rm gf}[\varphi]
\eeq
where $S_{\rm sym}[\varphi]$ is the gauge-invariant classical QED action and $S_{\rm gf}[\varphi]$ is the gauge-fixing term. We consider the linear gauge transformation of the fields
\beq
\label{eq:inftransfo}
 \varphi\to\varphi^{(\alpha)}={\cal A}\,\varphi+{\cal B}\quad
 {\rm and}\quad\cG\to\cG^{(\alpha)}={\cal A}\,\cG\,{\cal A}^t\,,
\eeq
where we use the notation ${\cal A}\,\varphi\equiv\int_y\sum_n{\cal A}_{mn}(x,y)\,\varphi_n(y)$. For the transformation \eqn{eq:gauge}, one has
${\cal A}_{mn}(x,y)=\delta_{mn}\delta^{(4)}(x-y)\exp[iq_m\alpha(x)]$ and ${\cal B}(x)=(-\p\alpha(x)/e,0,0)^t$. 

We show in \Ref{Reinosa:2007vi} that, for linear gauges, the 2PI effective action must be of the form
\beq
\label{eq:2PIsym}
 \Gamma_{\rm 2PI}[\varphi,\cG]=\Gamma_{\rm 2PI}^{\rm sym}[\varphi,\cG]+S_{\rm gf}[\varphi]\,,
\eeq
where $\Gamma_{\rm 2PI}^{\rm sym}[\varphi,\cG]$ is invariant under the gauge transformation \eqn{eq:inftransfo}. This generalizes the standard result that, for linear gauges, $S_{\rm gf}[\varphi]$ does not receive any loop corrections.

Equation \eqn{eq:2PIsym} encodes all the symmetry identities of the quantum theory. For instance, since the gauge-fixing term in \Eqn{eq:2PIsym} is $\cG$-independent, one concludes that the physical correlator $\bcG[\varphi]$ is obtained as the extremum of the symmetric functional $\Gamma_{\rm 2PI}^{\rm sym}[\varphi,\cG]$. It follows that it transforms covariantly under the gauge-transformation of its argument: $\bcG[\varphi^{(\alpha)}]=\bcG^{(\alpha)}[\varphi]$. For an infinitesimal transformation, this can be rewritten as (denoting by $\delta^{(\alpha)}$ the corresponding variations)
\beq
\label{eq:2PIWI2}
 \delta^{(\alpha)}\varphi_p\,\frac{\delta\bar\Sigma_{mn}[\varphi]}{\delta\varphi_p}=-\delta^{(\alpha)}\bcG^{-1}_{mn}[\varphi]\,.
\eeq
\Eqn{eq:2PIWI2} generates, through functional derivatives, a hierarchy of symmetry identities for the 2PI vertex functions \eqn{eq:npoint2}. It is remarkable that, despite the rather unusual definition of the latter, the obtained identities have the very same form as the usual WT identities. 
As an illustration, the ($\bar\psi,\psi$)-component of \Eqn{eq:2PIWI2} leads to the usual relation between three-point photon-fermion vertex function $\delta\bar\Sigma_{\bar\psi\psi}/\delta A^\mu|_{\bar\varphi}$ and the inverse fermion two-point function $i\bar D^{-1}$:
\beq
\label{eq:2PIWIex1}
 -\frac{1}{e}\,\p^\mu_z\left.\frac{\delta\bar\Sigma_{\bar\psi\psi}(x,y)}{\delta A^\mu(z)}\right|_{\bar\varphi}=\Big[\delta^{(4)}(x-z)-\delta^{(4)}(z-y)\Big]\,i\bar D^{-1}(x,y)\,.
\eeq
Similarly, one obtains from the $(A,A)$-component of \Eqn{eq:2PIWI2} that 2PI $n$-photon vertex functions with $\smash{n\ge3}$ are transverse in momentum space with respect to external momenta associated with field derivatives (see \cite{Reinosa:2007vi} for details):
\beq
\label{eq:2PIWIex2}
 \p^{\mu_1}_{z_1}\left.\frac{\delta\bar\Sigma_{AA}^{\rho\sigma}(x,y)}{\delta A^{\mu_1}(z_1)\cdots\delta A^{\mu_k}(z_k)}\right|_{\bar\varphi}=0\,.
\eeq

Next, using \Eqn{eq:stat}, one finds that the 2PI-resummed effective action reads
\beq
\label{eq:gammawi}
 \Gamma[\varphi]=\Gamma^{\rm sym}[\varphi]+S_{\rm gf}[\varphi]\,,
\eeq
where $\Gamma_{\rm sym}[\varphi]=\Gammatpi^{\rm sym}[\varphi,\bcG[\varphi]]$ is invariant under the gauge transformation \eqn{eq:inftransfo}. It immediately follows that 2PI-resummed vertices satisfy the standard WT identities. It is remarkable that, although they are a priori very different objects, 2PI and 2PI-resummed vertices with three external legs or more independently satisfy the same WT identities. 

An interesting byproduct of this analysis is that the 2PI two-point function itself, $i\bcG^{-1}\equiv i\bcG^{-1}[\bar\varphi]$, is not constrained by the underlying symmetry. This is because it is defined as the solution of a stationarity condition, \Eqn{eq:stat}, and not as a field derivative of some functional. In general, only the latter are constrained by symmetry identities. For instance, the 2PI photon polarization tensor $\bar\Pi^{\mu\nu}\equiv\bar\Sigma_{AA}^{\mu\nu}[\bar\varphi]$ is not constrained to be transverse in momentum space at any finite approximation order.\footnote{Of course a non-vanishing longitudinal component of $\bar\Pi$ is a pure artifact of the approximation. For instance, for a systematic loop expansion, it is always of higher order than the approximation order \cite{Reinosa:2006cm}.} In contrast, the 2PI-resummed two-point function
$\Gamma^{(2)}\equiv \delta^2\Gamma[\varphi]/\delta\varphi^2|_{\bar\varphi}$, being defined as a geometrical object, is constrained in the usual way by the gauge symmetry. In particular, the 2PI-resummed photon polarization tensor $\Pi_{\mu\nu}\equiv\delta^2(\Gamma[\varphi]-S_0[\varphi])/\delta A^\nu\delta A^\mu|_{\bar\varphi}$ is transverse in momentum space:
\beq
\label{eq:Pitransv}
 \p_x^\mu \Pi_{\mu\nu}(x,y)=0\,.
\eeq

It is important to realize that all the results derived above are direct consequences of the symmetry property \eqn{eq:2PIsym}. It follows that any 2PI approximation which respects this property leads to nonperturbative expressions for (2PI and 2PI-resummed) vertex functions which exactly satisfy (2PI) WT identities.
In \Ref{Reinosa:2007vi}, we give general rules to construct such approximation schemes for general abelian gauge theories. In particular, we show that, in QED, the 2PI loop-expansion satisfies the symmetry constraint \eqn{eq:2PIsym} at any approximation order.

\section{Renormalization \cite{Reinosa:2006cm,QED2}}

We define the rescaled field and propagator using  the matrix $Z\equiv {\rm diag}\,(Z_3,Z_2,Z_2)$:
\beq\label{eq:sfield_redef}
\varphi_R= Z^{-1/2}\varphi\qquad{\rm and}\qquad\cG_R= Z^{-1/2}\,\cG\,Z^{-1/2}\,,
\eeq
as well as the usual renormalized parameters $Z_0 m_R= Z_2 m$, $Z_1 e_R=Z_2Z_3^{1/2}e$ and $Z_4\lambda_R=Z_3\lambda$. The renormalized 2PI effective action, defined as $\smash{\Gamma_{\rm
2PI}^R[\varphi_R,\cG_R]=\Gamma_{\rm 2PI}[\varphi,\cG]}$, can be written as, up to a constant contribution:
\beq
\label{eq:GR2PI}
\Gamma_{\rm
2PI}^R[\varphi_R,\cG_R]=S_{0,R}[\varphi_R]+\frac{i}{2}\Str\ln\cG_R^{-1}+\frac{i}{2}\Str\,\cG_{0,R}^{-1}\,\cG_R+\Gammaint^R[\varphi_R,\cG_R]\,,
\eeq
where $S_{0,R}[\varphi_R]\equiv\frac{1}{2}\varphi_R^ti\cG_{0,R}^{-1}\,\varphi_R$
with $\cG_{0,R}^{-1}$ obtained from $\cG_{0}^{-1}$ after $(m,\lambda)$ is changed to $(m_R,\lambda_R)$. \Eqn{eq:GR2PI} defines $\Gammaint^R[\varphi_R,\cG_R]$, which can be expressed in terms of renormalized parameters and counterterms $\delta Z_i=Z_i-1$, $i=0,\ldots,4$:
\beq
\label{eq:deltaGammaint} \Gammaint^R[\varphi_R,\cG_R]=\Gammaint[\varphi_R,\cG_R;e_R]+\delta\Gammaint[\varphi_R,\cG_R]\,.
\eeq
The first term on the RHS is obtained from $\Gammaint[\varphi,\cG]$ by replacing bare quantities by renormalized ones whereas $\delta\Gammaint[\varphi_R,\cG_R]$ accounts for all counterterm contributions.
For a successful renormalization program, the latter should be such that the infinitely many UV-divergences appearing at any finite approximation order in the 2PI formalism can be canceled by adjusting a finite number of local counterterms consistent with the underlying symmetries, a non-trivial task obviously. 

In Refs.~\cite{Reinosa:2006cm,QED2}, we show that this can indeed be achieved if one includes in $\delta\Gammaint[\varphi_R,\cG_R]$ all counterterms contributions allowed by power counting and (Lorentz, gauge, etc.) symmetries. In the 2PI framework, this may allow for new type of counterterms, which have no analog in the standard renormalization theory, simply because there are more possibilities to construct symmetry invariants with both $\varphi_R$ and $\cG_R$ than with $\varphi_R$ alone. As an illustration, the most general counterterm contribution satisfying the above requirements at two-loop order can be written as \cite{QED2}:\footnote{Only \Eqn{eq:dGamma2app} receives higher-loop contributions.}
\beq
\label{eq:dGdecomp}
\delta\Gammaint[\varphi_R,\cG_R]=\delta S_{\rm int}[\varphi_R]+\delta\Gammaint^{\rm 1loop}[\varphi_R,\cG_R]+\delta\Gamma_2[\cG_R]\,,
\eeq
with
\bea\label{eq:Sintapp}
\delta S_{\rm int}[\varphi_R] & = & \int \!d^dx\, \frac{\delta Z_3}{2} A_R^{\mu}(x)\left(g_{\mu\nu}\partial_x^2-\partial^x_\mu\partial^x_\nu\right)A_R^\nu(x)\nn
&+& \int \!d^dx\,\,\bar\psi_R(x)\Big(i\delta Z_2\slashchar{\partial}_x-\delta
m-\delta Z_1e_R\,\slashchar{A}_R(x)\Big)\psi_R(x)\,,
\eea
\bea\label{eq:dGammaint1loop2app}
\delta\Gamma_{\rm int}^{\rm 1loop}[\varphi_R,\cG_R] & = &
\int \!d^dx\left(\frac{\delta\bar Z_3}{2}\left(g_{\mu\nu}\partial_x^2-\partial^x_\mu\partial^x_\nu\right)\!+\!\frac{\delta\bar M^2}{2}g_{\mu\nu}\!+\!\frac{\delta\bar\lambda}{2}\partial^x_\mu\partial^x_\nu\right) G_R^{\mu\nu}(x,y)\Big|_{y=x}\nn
&-&\int \!d^dx\,\,\tr\,\Big(i\delta \bar Z_2\slashchar{\partial}_x-\delta\bar m-\delta \bar Z_1e_R\,\slashchar{A}_R(x)\Big) D_R(x,y)\Big|_{y=x}\nn
&-&\int \!d^dx\,\,\delta\tilde Z_1 e_R\Big(\bar\psi_R(x)\slashchar{K}_R(x,x)+\bar{\slashchar{K}}_R(x,x)\psi_R(x)\Big)\,,
\eea
and
\beq\label{eq:dGamma2app}
\delta\Gamma_2[\cG_R]=\int \!d^dx\,\left(\frac{\delta\bar g_1}{8}\,G_{R\,\mu}^\mu(x,x)G_{R\nu}^\nu(x,x)+\frac{\delta\bar
g_2}{4}\,G_R^{\mu\nu}(x,x)G^R_{\mu\nu}(x,x)\right)\,.
\eeq
A few remarks are in order here. First, notice that some terms, which would be allowed by power counting and global symmetries, such as $A_R^2$, or $(\partial A_R)^2$ in \Eqn{eq:Sintapp}, or $A_R^2 G_R$ in \Eqn{eq:dGammaint1loop2app}, are forbidden by the gauge symmetry. Correspondingly, as shown in \cite{QED2}, 2PI WT identities prevent the appearance of those UV divergences which would call for such counterterms. In contrast, since the photon correlator $G_R$ is invariant under the gauge transformation \eqn{eq:inftransfo} terms of the form $G_R$, $\partial\partial G_R$ and $G_R^2$ are allowed, giving rise to corresponding counterterms $\delta\bar M^2$, $\delta\bar\lambda$, $\delta\bar g_1$ and $\delta\bar g_2$, which would be absent in the exact theory. As discussed in detail in \cite{Reinosa:2006cm} these counterterms actually serve to absorb divergences in the longitudinal (in momentum space) part of the inverse photon correlator $\bar G_R^{-1}$ which, as discussed previously, are not forbidden by 2PI WT identities. Notice though that, as already mentioned, such divergences are a pure artifact of the approximation. For a 2PI loop-expansion, they are systematically of higher order than the approximation order. So are the (divergent part of the) corresponding counterterms which, therefore, vanish as one approaches the exact theory. 

Next, notice that gauge symmetry allows for a priori different independent $\delta Z$'s, such as $\delta Z_3$ and $\delta\bar Z_3$, etc. This is a generic feature of 2PI renormalization theory, related to the existence of a priori different possible definitions of vertex functions. The latter come with a priori different UV divergences, to be absorbed in the different $\delta Z$'s. Again, at a given order in the 2PI loop-expansion, these differences are systematically higher-order effects. Finally, the terms $\bar\psi_R\slashchar{\partial}\psi_R$, $\bar\psi_R\slashchar{A}_R\psi_R$, $\slashchar{\partial}D_R$ and $\slashchar{A}_RD_R$ are obviously related by the gauge symmetry and so are the counterterms $\delta Z_1$, $\delta Z_2$, $\delta \bar Z_1$ and $\delta\bar Z_2$. For instance, for a suitable definition of the renormalized charge $e_R$, one has $\delta Z_1=\delta Z_2$ and $\delta \bar Z_1=\delta\bar Z_2$. 

In Refs.~\cite{Reinosa:2006cm,QED2} we explicitly show that 2PI WT identities constrain the UV divergences appearing at any finite approximation order in such a way that they can all be absorbed in the set of counterterms discussed here.

We stress again that the profusion of counterterms encountered here is a generic feature of 2PI renormalization theory and is a mere artifact of the finite approximation: different counterterms are nothing but different approximations of the true counterterms of the theory. For instance, the difference $\delta Z_3-\delta\bar Z_3$, or the gauge-fixing-parameter counterterm $\delta\bar Z_4=\delta\bar\lambda/\lambda_R$ should approach their exact (zero) value as the approximation order increases. As mentioned above, this is certainly true for the divergent parts of the respective counterterms. To ensure that this be true for the finite parts as well, one needs to impose suitable renormalization conditions. A related issue is the fact that, for a successful  renormalization procedure, one needs to fix all the above counterterms without introducing any new input parameter -- other than $m_R$ and $e_R$ -- to specify the theory.

In the 2PI formalism, the conditions which fix the various counterterms can be separated in two distinct classes. The first ones are the usual renormalization conditions, which define the theory. These can, for instance, be written in terms of the 2PI-resummed two- and three-point vertices $\Gamma_R^{(2)}$ and $\Gamma_{R}^{(3)}$. To be more specific, we denote by $\Sigma_R^{\bar\psi\psi}$ and $\Sigma_{R}^{AA}\equiv\Pi_R$ the 2PI-resummed fermion and photon self-energies, i.e. the non-vanishing components of $\Sigma_R=i\Gamma_R^{(2)}+\cG_{0,R}^{-1}$, and by $\Gamma_R^{(2,1)}\equiv\Gamma_{R,A\psi\bar\psi}^{(3)}$ the fermion-photon three-point vertex. QED can be defined\footnote{To be complete, this should be supplemented by renormalization conditions for composite operators.} by the following set of independent renormalization conditions at a given renormalization point $*$ (chosen here to correspond to zero photon and on-shell fermion momenta\footnote{For a discussion of more generic renormalization points, see \cite{QED2}}):
\beq\label{eq:rencharge}
 \Sigma_R^{\bar\psi\psi}\Big|_*=0\quad,\quad\left.\frac{\p\Sigma_R^{\bar\psi\psi}}{\p \slashchar{p}}\right|_{*}=0\quad,\quad\left.\frac{d\Pi_{R}^T}{dk^2}\right|_{*}=0\quad{\rm and}\quad\left.\Gamma_{R,\mu}^{(2,1)}\right|_*=-e_R\gamma_\mu\,,
\eeq
where $\Pi_R^T$ is defined as, see \eqn{eq:Pitransv}, $\Pi_{R}^{\mu\nu}(k)=(g^{\mu\nu}-k^\mu k^\nu/k^2)\Pi_R^T(k^2)$. 

The second class of conditions are in fact consistency conditions, whose role is to restore the identity of the various vertex functions at the renormalization point. It is a necessary requirement in order to guarantee that the approximation scheme converges toward the correct theory. For instance, we demand that
the 2PI fermion self energy satisfies
\beq
\label{eq:consfermion}
 \bar\Sigma_R^{\bar\psi\psi}\Big|_*=\Sigma_R^{\bar\psi\psi}\Big|_*\quad{\rm and}\quad
 \left.\frac{\p\bar\Sigma_R^{\bar\psi\psi}}{\p \slashchar{p}}\right|_{*}=\left.\frac{\p\Sigma_R^{\bar\psi\psi}}{\p \slashchar{p}}\right|_{*}\,.
\eeq
Similarly, we impose, for the longitudinal and transverse part of the 2PI photon polarization tensor,
\beq
\label{eq:2PIresrenG} \,
 \bar\Pi_{R}^L\Big|_*=0\quad, \quad\left.\frac{d\bar\Pi_{R}^L}{dk^2}\right|_{*}=0\quad{\rm and}\quad\left.\frac{d\bar\Pi_{R}^T}{dk^2}\right|_{*}=\left.\frac{d\Pi_{R}^T}{dk^2}\right|_{*}\,.
\eeq
where the first two conditions follow from the fact that, at any approximation order, the 2PI-resummed photon polarization tensor $\Pi_R$ is transverse in momentum space, i.e. $\Pi_R^L(k^2)=0$, see \eqn{eq:Pitransv}. Similar conditions for the 2PI four-photon function allows one to fix the four-photon counterterms $\delta \bar g_1$ and $\delta \bar g_2$ \cite{Reinosa:2006cm}. Finally, we demand that the different definitions of the three-point vertex coincide at the renormalization point:\footnote{Lorentz symmetry and charge-conjugation invariance imply that $\smash{\delta\bar\Sigma_R^{\bar\psi A}/\delta \psi_R|_*=\delta\bar\Sigma_R^{A\psi}/\delta\bar\psi_R|_*}$ \cite{QED2}.}
\beq\label{eq:renvertices}
 \left.\frac{\delta\bar\Sigma_R^{\bar\psi\psi}}{\delta A_R}\right|_*=\left.\frac{\delta\bar\Sigma_R^{\bar\psi A}}{\delta \psi_R}\right|_*=\Gamma_{R}^{(2,1)}\Big|_*\,,
\eeq
In total, the conditions \eqn{eq:rencharge}-\eqn{eq:renvertices} allow one to fix all the counterterms introduced above. We stress that conditions \eqn{eq:consfermion}-\eqn{eq:renvertices} are imposed by the consistency of the approximation scheme. The only freedom one has lies in the four independent renormalization conditions \eqn{eq:rencharge}, as expected for QED.

\section{Conclusion and Outlook}

To summarize, we have proposed a consistent, gauge-invariant 2PI renormalization procedure for abelian gauge theories and have clarified a number of issues concerning (gauge) symmetries in the 2PI formalism. We believe this is an important step, which opens the way to reliable quantitative calculations in gauge theories using nonperturbative 2PI techniques. Possible applications include equilibrium and nonequilibrium dynamics. This work is also an important step toward the case of non-abelian gauge theories \cite{Calzetta:2004sh,Berges:2004pu}.

Here, we would like to comment about another important issue concerning gauge theories in the 2PI formalism, namely the possibility (or not) to define gauge-fixing independent quantities. The point is that physical observables computed from the 2PI effective action usually contain residual, spurious gauge-fixing dependences at finite approximation order\footnote{The fact that the (renormalized) 2PI-resummed effective action is gauge-invariant does not guarantee that observables obtained from it are gauge-fixing independent.}, see e.g.~\cite{Arrizabalaga:2002hn,Mottola:2003vx}. No systematic procedure has been found so far to get rid of the latter. General results~\cite{Arrizabalaga:2002hn} show however that these gauge dependent contributions are parametrically suppressed in powers of the coupling, which indicates that they should at least be well under control at weak coupling. Moreover, the observed good convergence properties of 2PI approximations schemes, see e.g. \cite{Berges:2004hn}, suggest that gauge-fixing dependences may be controlled beyond the perturbative region. This has recently been tested for QED in \Ref{Borsanyi:2007bf}, where the thermodynamic pressure has been computed from the (renormalized) 2PI loop-expansion at two-loop order in the covariant gauge. The results indicate that gauge-fixing parameter dependences remain under control in a wide range of couplings and are comparable with renormalization scale dependences, another source of uncertainty inherent in such calculations. Moreover, the Landau gauge has been identified as the gauge minimizing both gauge-fixing parameter and renormalization scheme dependences.

\end{document}